# Traveling Surface Spin-Wave Resonance Spectroscopy Using Surface Acoustic Waves

P. G. Gowtham[1], T. Moriyama[2], D. C. Ralph,[1,3] and R. A. Buhrman[1]

[1]Cornell University, Ithaca, New York, 14853, USA

[2]Institute for Chemical Research, Kyoto University, Kyoto, Japan

[3]Kavli Institute at Cornell, Ithaca, New York, 14853, USA

## Abstract

Coherent gigahertz-frequency surface acoustic waves (SAWs) traveling on the surface of a piezoelectric crystal can, via the magnetoelastic interaction, resonantly excite traveling surface spin waves in an adjacent thin-film ferromagnet. These excited surface spin waves, traveling with a definite in-plane wave-vector $\mathbf{q}_\parallel$ enforced by the SAW, can be detected by measuring changes in the electro-acoustical transmission of a SAW delay line. Here, we provide a demonstration that such measurements constitute a precise and quantitative technique for spin-wave spectroscopy, providing a means to determine both isotropic and anisotropic contributions to the spin-wave dispersion and damping. We demonstrate the effectiveness of this spectroscopic technique by measuring the spin-wave properties of a Ni thin film for a large range of wave vectors, $|\mathbf{q}_\parallel| = 2.5 \times 10^4 - 8 \times 10^4$ cm$^{-1}$, over which anisotropic dipolar interactions vary from being negligible to quite significant.

## I. INTRODUCTION

Spin waves in magnetic materials can transport spin information with a high degree of efficiency over distances that far exceed the limitations of spin diffusion in metals.[1–3] Consequently, control over the generation and propagation of spin waves in micron-scale and nano-scale structures is of interest for next-generation spin-based technologies. Considerable thought and effort has been brought to bear on exploiting systems with novel anisotropic spin-spin interactions (e.g. Dzyaloshinskii-Moriya interactions)[4–10] and on engineering new structures (e.g. magnonic crystals) for use in tailoring the propagation characteristics of spin waves[11–18]. These systems can possess highly anisotropic spin-wave dispersion and damping – both of which can be used to advantage for guiding and manipulating spin waves in magnetic heterostructures. Developing a spectroscopic technique that is capable of quantitatively measuring such anisotropies (and on length scales that are technologically relevant) is therefore imperative.

Recent experiments[19–22] have shown that surface acoustic wave (SAW) delay line devices can (via the magnetoelastic interaction) be used to launch and detect spin waves in magnetic thin films that are coupled to piezoelectric substrates. Here we extend these initial results to show that the SAW-based excitation of traveling surface spin waves provides a sensitive spectroscopic technique for making quantitative measurements of anisotropic contributions to the spin-wave dispersion and damping. Unlike spin-wave measurement techniques which possess no wave-vector selectivity (*e.g*, anomalous Nernst effect[23], and spin-pumping[24–26]/inverse spin Hall effect detection schemes[1,27–29]), a SAW can excite a single traveling surface spin wave mode with a definite in-plane wave vector $\mathbf{q}_\parallel$ that is matched to the wave vector chosen for the SAW. While other acoustical techniques (*e.g.,* bulk opto-acoustical techniques) for spin-wave spectroscopy can be used to spectrally select and measure single spin-wave modes, these techniques suffer from the difficulty that quantitative analysis of the spin-wave amplitude line-shapes and spin-wave resonance frequencies can be very challenging[30] – a difficulty that we will show SAW-based traveling surface spin-wave spectroscopy does not share. The SAW delay line measurement scheme differs from other electrical spin-wave spectroscopy techniques (*e.g.* microstrip delay lines[31,32]) in that the SAW imposes an effective pump field modulated at wave-vector $\mathbf{q}_{\|}$ throughout the entire magnetic film that drives the traveling spin wave and therefore provides a fairly direct measure of the dynamic, wave-vector dependent spin-wave susceptibility – whereas microstrip delay line techniques provide a measure of the local spin-wave excitation/non-local propagator (i.e. Green's function) for a magnetic film. The low velocity of a Rayleigh SAW implies that at the GHz frequency scales characteristic of spin-wave resonance, the wave-vector $\mathbf{q}_{\|}$ of the effective pump field and traveling spin-wave can span characteristic scales ($\mathbf{q}_{\|} \sim 10^4 – 10^6$ cm$^{-1}$) over which various isotropic and anisotropic spin-spin interactions start to become important in determining the propagation of the spin-wave. While operable only at discrete values of $|\mathbf{q}_{\|}|$ and requiring both spin-wave excitation *and* detection, unlike wave-vector resolved optical spin-wave measurement schemes such as Brillouin light scattering (BLS)[12,33–36], SAW-driven spin wave spectroscopy can in principle be operated at wave-vectors (*e.g.*, $|\mathbf{q}_{\|}| > 3\times10^5$ cm$^{-1}$), currently inaccessible to BLS, that extend deep into the dipole-exchange regime – as will be discussed more thoroughly in the concluding Outlook Section.

SAW-based spin-wave spectroscopy allows for quantitative studies of the traveling surface spin-wave susceptibility, dispersion, and damping as a function of varying the angle between the orientation of the magnetization $\mathbf{m_0}$ and fixed $\mathbf{q}_{\|}$. Such analysis as a function of $|\mathbf{q}_{\|}|$ and angle provides a simple means to directly determine anisotropic and wave-vector-

dependent contributions (*e.g.*, from anisotropic spin-spin interactions) to the spin-wave dispersion and damping. To demonstrate these capabilities, we perform SAW-driven spin-wave resonance measurements as a function of applied field and in-plane field angle for a $d = 10$ nm thick Ni thin film microstrip on a piezoelectric YZ-cut LiNbO$_3$ substrate. We examine a range of larger wave vector, $|\mathbf{q}_\parallel| \sim 2.5\times10^4$ to $8\times10^4$ cm$^{-1}$, than has been studied previously by SAW experiments. This range is chosen to span from the region where anisotropic dipolar interactions should be negligible to the region where they contribute significantly to the spin-wave dispersion. By implementing a quantitative analysis of absorption measurements for SAW delay lines, we demonstrate that it is possible to achieve a comprehensive, quantitative determination of the wave-vector and angular structure of this dipolar interaction. The same experimental technique should also be more generally applicable to measure other interactions that modify spin-wave propagation, *e.g.* in magnonic crystals and magnetic heterostructures designed to have strong Dzyaloshinskii-Moriya interactions.

## II. ANALYTICAL CALCULATION OF SAW POWER ABSORBED BY EXCITATION OF A SPIN WAVE

The spin-wave dynamics of an ultrathin magnetic film driven coherently by a SAW traveling with an in-plane wave vector $\mathbf{q}_\parallel$, and the resultant SAW power absorption, can be derived within the framework of the Landau-Lifshitz-Gilbert (LLG) equation[37]

$$\frac{d\mathbf{m}(\mathbf{r})}{dt} = -\gamma \mathbf{m}(\mathbf{r}) \times \mathbf{H}_{\text{eff}}(\mathbf{r}) + \Gamma(\mathbf{q}_\parallel, \mathbf{m_0})\mathbf{m}(\mathbf{r}) \times \frac{d\mathbf{m}(\mathbf{r})}{dt}, \quad (1)$$

where $\mathbf{m}(\mathbf{r})$ is the local magnetization, $\mathbf{H}_{\text{eff}}(\mathbf{r})$ is the local effective field exerting a torque on $\mathbf{m}(\mathbf{r})$, and $\mathbf{m_0}$ is the equilibrium magnetization in the absence of the RF pump field. We allow for the *spin-wave* damping term $\Gamma(\mathbf{q}_\parallel, \mathbf{m_0})$ to depend on $\mathbf{q}_\parallel$ and $\mathbf{m_0}$ rather than assuming it is a constant, as is sometimes done, but we do employ the assumption $\Gamma(\mathbf{q}_\parallel, \mathbf{m_0}) \ll 1$ so that the damping term can be treated as a perturbation. Eqn. (1) is analyzed using both a $\zeta\eta\xi$ and a $xyz$ coordinate system (Figure 1). The $\zeta\eta\xi$ coordinate system is defined such that $\mathbf{q}_\parallel$ of the SAW lies along the $+\eta$ direction, the $\zeta$ axis corresponds to the magnetic easy axis of our thin Ni microstrip (which is in-plane), and the $\xi$ axis lies normal to the film plane. The $xyz$ coordinate system, which is defined such that $\mathbf{m_0}$ lies along the $z$ axis, is convenient for deriving the linearized LLG spin-wave dynamics about equilibrium.

The various components of the effective field relevant to a continuous Ni microstrip are

$$\mathbf{H}_{\text{eff}} = \mathbf{H}_{\text{app}} + \mathbf{H}_{\mathbf{k}} + \mathbf{H}_{\text{anis}}^{\perp} + \mathbf{h}_{\text{RF}}(\mathbf{r},t) + \frac{2A_{ex}}{M_s}\nabla^2\mathbf{m}(\mathbf{r},t) + \mathbf{h}_{\mathbf{d}}(\mathbf{r},t). \tag{2}$$

$\mathbf{H}_{\text{app}}$ is the external field applied in the plane of the film. $\vartheta_H$ is defined as the angle $\mathbf{H}_{\text{app}}$ makes with respect to the $\eta$ axis (see Figure 1). The internal contributions to $\mathbf{H}_{\text{eff}}$ are the in-plane anisotropy field $\mathbf{H}_{\mathbf{k}} = H_k m_\zeta \hat{\boldsymbol{\zeta}}$, a perpendicular anisotropy field $\mathbf{H}_{\text{anis}}^{\perp} = \frac{2K_\perp}{M_s}m_\xi\hat{\boldsymbol{\xi}}$ that partially counteracts the out-of-plane demagnetization field, the magnetoelastic pump field $\mathbf{h}_{\text{RF}}(\mathbf{r},t)$ at wave vector $\mathbf{q}_\parallel$ generated by the SAW via the magnetoelastic interaction, the exchange field $\frac{2A_{ex}}{M_s}\nabla^2\mathbf{m}(\mathbf{r},t)$ where $A_{ex}$ is the exchange stiffness, and finally the dipolar field $\mathbf{h}_{\mathbf{d}}(\mathbf{r},t)$ that encompasses both the out-of-plane demagnetization energy as well as a fluctuating, spatially varying component generated by the temporal and spatial variation of the magnetization in the spin wave. This non-local dipolar interaction serves to couple various parts of the spin-wave together and thus affects the propagation of the spin-wave. The non-local fluctuating part of the dipolar interaction has not been included in previous studies analyzing SAW-driven spin-wave resonance in magnetic thin films.[21,22]

The in-plane equilibrium magnetic orientation $\mathbf{m_0}$ is completely determined by $\mathbf{H}_{app}$ and $\mathbf{H}_k$. We transform Eqns. (1) and (2) into the $xyz$ coordinate system and linearize the LLG equation about $\mathbf{m_0}$. The magnetoelastic pump field $\mathbf{h}_{RF}(\mathbf{r},t) = \mathbf{h}_{RF}^{\mathbf{q}_\parallel} e^{i(\mathbf{q}_\parallel \cdot \mathbf{r} - \omega t)}$ arising from the traveling SAW derives from the relation $\mathbf{h}_{RF}(\mathbf{r},t) = -\frac{1}{M_s}\nabla_\mathbf{m} f_{ME}$ where $f_{ME}$ is the magnetoelastic part of the magnetic free energy density

$$f_{ME} = -B_{eff}\varepsilon_{\eta\eta}e^{i(\mathbf{q}\cdot\mathbf{r}-\omega t)}\sin^2\theta\cos^2\varphi + B_{shear}\varepsilon_{\eta\xi}e^{i(\mathbf{q}\cdot\mathbf{r}-\omega t)}\cos\theta\sin\theta\sin\varphi \qquad (3)$$
$$+ B_{\xi\xi}\varepsilon_{\xi\xi}e^{i(\mathbf{q}\cdot\mathbf{r}-\omega t)}\cos^2\theta.$$

The angle $\varphi$ denotes the in-plane azimuthal angle that $\mathbf{m}$ makes with the $\zeta$ axis, $\theta$ is the magnetization polar angle defined with respect to the $\xi$ axis, and $B_{eff}$, $B_{shear}$, and $B_{\xi\xi}$ are respectively the effective magnetoelastic couplings to the in-plane longitudinal strain amplitude $\varepsilon_{\eta\eta}$, shear strain amplitude $\varepsilon_{\eta\xi}$ and strain amplitude perpendicular to the Ni film plane associated with the traveling SAW. Since $\mathbf{m_0}$ in our measurements is in-plane, we ignore the component of $f_{ME}$ that is sensitive to $\varepsilon_{\xi\xi}$. We also ignore the shear strain $\varepsilon_{\eta\xi}$ as the Ni film under study is a $d = 10$ nm thick film and is in the regime where $d \ll \lambda_{SAW}$ for all the SAW bandpasses employed in this experiment. In this thin-film regime, the zero-shear-strain boundary condition at the substrate surface (which the SAW much satisfy) implies that the shear strain $\varepsilon_{\eta\xi}$ through the film is small. Then $\mathbf{h}_{\mathbf{RF}}^{\mathbf{q}_\parallel}$ can be expressed in the $xyz$ coordinate system as

$$\mathbf{h}_{\mathbf{RF}}^{\mathbf{q}_\parallel} = +\frac{2B_{eff}}{M_s}\varepsilon_{\eta\eta}\sin(\varphi_0)\cos(\varphi_0)\hat{\mathbf{x}}. \tag{4}$$

where $\varphi_0$ is the angle that $\mathbf{m}_0$ makes with the $\zeta$ (easy) axis, and $M_s$ is the saturation magnetization.

We solve for the traveling spin-wave amplitude $\delta\mathbf{m}^{\mathbf{q}_\parallel}$ (averaged over the film thickness) in terms of $\mathbf{h}_{\mathbf{RF}}^{\mathbf{q}_\parallel}$ by self consistently solving both the LLG equations linearized about $\mathbf{m}_0$ for the dynamic magnetization profile across the Ni film thickness (Eqns. (5) and (6)) as well as the magnetostatic equations (Eqns. (5) and (6)) for the nonlocal dipolar fields $\mathbf{h}_d(\mathbf{r},t)$ that depend on the instantaneous magnetization profile of the traveling spin-wave.

$$-i\omega\delta m_x^{\mathbf{q}_\parallel}(y) = -\gamma\left(H_k\cos^2(\varphi_0) + H_{app}\sin(\varphi_0 + \vartheta_H) + \frac{2A_{ex}}{M_s}\left[|\mathbf{q}_\parallel|^2 - \frac{\partial^2}{\partial y^2}\right]\right)\delta m_y^{\mathbf{q}_\parallel}(y) \tag{5}$$
$$+\gamma h_d^y(y) + \gamma h_{RF,y}^{\mathbf{q}_\parallel} + i\omega\Gamma(\mathbf{q}_\parallel,\mathbf{m}_0)\delta m_y^{\mathbf{q}_\parallel}(y)$$

$$-i\omega\delta m_y^{\mathbf{q}_\parallel}(y) = +\gamma\left(H_k\cos(2\varphi_0) + H_{app}\sin(\varphi_0 + \vartheta_H) + \frac{2A_{ex}}{M_s}\left[|\mathbf{q}_\parallel|^2 - \frac{\partial^2}{\partial y^2}\right]\right)\delta m_x^{\mathbf{q}_\parallel}(y) \tag{6}$$
$$-\gamma h_d^x(y) - \gamma h_{RF,x}^{\mathbf{q}_\parallel} - i\omega\Gamma(\mathbf{q}_\parallel,\mathbf{m}_0)\delta m_x^{\mathbf{q}_\parallel}(y)$$

$$\nabla\cdot\mathbf{h}_d = -4\pi\nabla\cdot\mathbf{m} \tag{7}$$

$$\nabla\times\mathbf{h}_d = 0 \tag{8}$$

This enables us to express $\mathbf{h}_d$ in terms of magnetostatic potential $\Phi$ as $\mathbf{h}_d = -\nabla\Phi$ and then solve for the potential $\Phi$ and $\delta\mathbf{m}^{\mathbf{q}_\parallel}(y)$. The thickness dependence of the fluctuating component $\delta \mathbf{m}^{\mathbf{q}_{\|}}(y)$ of the magnetization is non-zero and arises from considering the evanescence of the traveling SAW into the Ni film (with decay length of order $\lambda_{SAW}$) as well as boundary conditions on the magnetostatic potential $\Phi$ associated with $\mathbf{h}_\mathbf{d}(\mathbf{r},t)$ at the surfaces of the magnetic film. The calculation is adapted from Stamps and Hillebrands[38] and further details of the solution can be found there.

The relationship between the thickness averaged spin-wave amplitude and the pump field is expressed as $\delta \mathbf{m}^{\mathbf{q}_{\|}} = \overline{\overline{\chi}} \cdot \mathbf{h}_{\mathbf{RF}}^{\mathbf{q}_{\|}}$ where $\overline{\overline{\chi}} = \overline{\overline{\chi}}' + i\overline{\overline{\chi}}''$ is the susceptibility tensor. We have restricted ourselves to the condition $|\mathbf{q}_{\|}|d \ll 1$, where $d$ is the film thickness, appropriate for the wave-vector range $|\mathbf{q}_{\|}| = 2.5 \times 10^4 - 8 \times 10^4$ cm$^{-1}$ and $d = 10$ nm thin film microstrips employed in this study. The imaginary part of the susceptibility governing the out-of-phase response of the magnetization to the relevant component of the magnetoelastic pump field is then

$$[\overline{\overline{\chi}}'']_{xx} = \frac{\frac{\omega_p}{\gamma}\Gamma(\mathbf{q}_{\|},\mathbf{m_0})\left(\Upsilon^2 + \left(\frac{\omega_p}{\gamma}\right)^2\right)}{\left(\left(\frac{\omega_{res}}{\gamma}\right)^2 - \left(\frac{\omega_p}{\gamma}\right)^2\right)^2 + \left(\frac{\omega_p \Gamma(\mathbf{q}_{\|},\mathbf{m_0})(\Psi+\Upsilon)}{\gamma}\right)^2}, \quad (9)$$

where $\omega_p = c_{SAW}|\mathbf{q}_{\|}|$ is the fixed SAW pump frequency, $c_{SAW}$ is the Rayleigh SAW sound speed, and $\omega_{res} = \gamma\sqrt{\Psi\Upsilon}$ can be identified as the traveling surface dipole-exchange spin-wave resonance frequency, and the quantities $\Psi$ and $\Upsilon$ are

$$\Psi = H_k \cos(2\varphi_0) + H_{app}\sin(\varphi_0 + \vartheta_H) + \frac{2A_{ex}}{M_s}|\mathbf{q}_{\|}|^2 + 2\pi M_s |\mathbf{q}_{\|}|d\cos^2\varphi_0$$
$$\Upsilon = H_k \cos^2\varphi_0 + H_{app}\sin(\varphi_0 + \vartheta_H) + \frac{2A_{ex}}{M_s}|\mathbf{q}_{\|}|^2 + \left(4\pi M_s - \frac{2K_\perp}{M_s}\right) - 4\pi M_s\left(\frac{|\mathbf{q}_{\|}|d}{2}\right). \quad (10)$$

The SAW power absorbed due to the excitation of a traveling surface dipole-exchange spin wave at wave vector $\mathbf{q}_\parallel$, using the relation $P_{abs} = \frac{\omega_p}{2} \mathbf{h}_{RF}^{\mathbf{q}_\parallel \dagger} \cdot \overline{\chi}'' \cdot \mathbf{h}_{RF}^{\mathbf{q}_\parallel}$ [21,39], is

$$P_{abs} = \frac{\omega_p}{2} \left[\overline{\chi}''\right]_{xx} \left(\frac{2B_{eff}}{M_s} \varepsilon_{\eta\eta}\right)^2 \sin^2 \varphi_0 \cos^2 \varphi_0. \tag{11}$$

The angular structure of the absorbed SAW power derives from a combination of the magnetoelastic RF field contribution $\left|\mathbf{h}_{RF}^{\mathbf{q}_\parallel}\right|^2 = \left(\frac{2B_{eff}}{M_s} \varepsilon_{\eta\eta}\right)^2 \sin^2 \varphi_0 \cos^2 \varphi_0$ and the susceptibility $\left[\overline{\chi}''\right]_{xx}$. The pump field itself depends on the angle that $\mathbf{m}_0$ makes with respect to the $\eta$ axis (the longitudinal strain axis) and possesses four-fold symmetry in $\varphi_0$ with maxima when $\varphi_0 = (2n+1)\pi/4$ for $n \in \mathbb{Z}$. This magnetoelastic pump component has the same form for any spin wave within an in-plane magnetized polycrystalline thin film excited by a coherent Rayleigh SAW. It is the $\left[\overline{\chi}''\right]_{xx}$ component of $P_{abs}$ that carries both the information about the internal magnetic anisotropy energies present in a specific system and also, more importantly, the angular and wave-vector dependence of the excited spin wave. An inspection of Eqns. (9), (10) and (11) shows that delay line measurements of $P_{abs}$ as a function of $\varphi_0$ and $|\mathbf{q}_\parallel|$ can be used to determine both the isotropic and anisotropic parts of the spin-spin interactions embedded within $\left[\overline{\chi}''\right]_{xx}$, given an independent determination of $M_s$, $H_k$, and $K_\perp$. For our continuous Ni film, the *isotropic* component of the spin-wave corrections comes from (1) the exchange interaction and (2) the term $4\pi M_s(|\mathbf{q}_\parallel|d/2)$ in $\Upsilon$ (Eqn. (10)) whose origin is the dipolar interaction. The *anisotropic* component in our Ni films is expected to arise solely from the dipolar interaction and is completely encoded in the term $2\pi M_s |\mathbf{q}_\parallel| d \cos^2 \varphi_0$ within $\Psi$. We show in the remainder of the paper that SAW based spectroscopy can sensitively and coherently map out the $\varphi_0$ and $|\mathbf{q}_\parallel|$ dependence arising from the dipolar interaction even in the low $M_s$ Ni system at moderate wave-vectors.

### III. EXPERIMENT

We performed SAW power absorption measurements on Al(10 nm)/AlO$_x$(2 nm)/Ni(10 nm)/Pt(15 nm) microstrips defined in the middle of a SAW delay line. The Al/AlO$_x$ underlayer and Pt overlayer were deposited to enable a separate inverse spin Hall effect (ISHE) quantification of the SAW-induced spin-wave excitation (not discussed in this paper). The film stack was magnetron sputter-deposited on 0.5-mm-thick single-side polished YZ-cut LiNbO$_3$ substrates and patterned via lift-off process into 100 μm × 500 μm microstrips with the long axis orthogonal to the LiNbO$_3$ Z-axis SAW propagation direction. The crystal Z-axis thus corresponds to the $\eta$ axis and the long axis of the wire corresponds to the $\zeta$ axis. The base pressure for our chamber was $P_0 < 4 \times 10^{-9}$ Torr and the working Ar gas pressure was kept at 2 mTorr throughout the deposition process. The samples were deposited with no applied magnetic field. The AlO$_x$ layer was formed by native oxidation in air. Al(100 nm) was deposited and patterned via lift-off for the delay line metallization. Our SAW delay line was designed to have a center frequency $f_0$ = 300 MHz and we use higher order bandpasses at $f_{pump}$ = 1.48, 2.67, 3.26, 3.86, and 4.45 GHz to excite magnetization dynamics in the Ni. Details about our SAW delay line are provided in Figure 2.

The SAW resonant absorption was determined as a function of $H_{app}$ and $\vartheta_H$ by measuring the quantity $\left|S_{21}^{loss}(H_{app},\vartheta_H)\right| = \left|S_{21}(H_{app},\vartheta_H)\right| - \left|S_{21}(H_{app}=3\text{ kOe},\vartheta_H)\right|$ at selected bandpasses of the SAW delay line using a vector network analyzer (VNA). At the frequencies employed in our experiment, when $H_{app} = 3$ kOe the magnetic system is far from any spin-wave resonance condition. In such case the loss is determined by changes in the transmission line impedance due to mass-loading and capacitive coupling to the magnetic metallic film. Figure 3 shows a density plot of $\left|S_{21}^{loss}(H_{app},\vartheta_H)\right|$ at the various bandpasses of our SAW delay line. The VNA transmission measurements were carried out using an input RF power of +10 dBm. The $(H_{app},\vartheta_H)$ dependence of the $\left|S_{21}^{loss}(H_{app},\vartheta_H)\right|$ as a function of $f_{pump}$ are consistent with the fact that the SAW is driving a magnetic resonance.

Calculated density plots of $P_{abs}(H_{app},\vartheta_H)$ as a function of the various $f_{pump}$ employed in the experiment are shown in the bottom row of Figure 3. In performing these calculations we used an in-plane anisotropy $H_k \sim 380 \pm 10$ Oe with easy axis along the $\zeta$ axis as measured by Anisotropic Magnetoresistance (AMR) measurements, $M_s = 485$ emu/cm$^3$ as measured by SQUID magnetometry, and $K_\perp \sim 1.13 \times 10^6$ ergs/cm$^3$ as measured by out-of-plane SQUID scans for inputs into $P_{abs}(H_{app},\vartheta_H)$ of Eqn. (11). SQUID and AMR transport characterization of the Ni films are shown in Figure 4. Both the large $H_k$ and substantial $K_\perp$ in our 10-nm thick Ni film are likely due to the magnetoelastic interaction and high anisotropic strains arising from depositing the film stack on the LiNbO3 substrate. We assumed a value $A_{ex} = 8 \times 10^{-7}$ erg/cm for the exchange stiffness of Ni[40], a speed $c_{SAW} = 3.488 \times 10^5$ cm/s for SAW propagation along the Z-axis of YZ-LiNbO$_3$[41], and $d = 10$ nm for the Ni film thickness. The only remaining undetermined quantity in the formula for $P_{abs}(H_{app}, \vartheta_H)$ (Eqn. (11)) is the spin-wave damping $\Gamma(\mathbf{q}_\parallel, \mathbf{m_0})$. The measured log scale $|S_{21}^{loss}(H_{app}, \vartheta_H)|$ was converted to linear scale and normalized between -1 and 0.

We find good quantitative agreement between this normalized linear $|S_{21}^{norm}(H_{app}, \vartheta_H)|$ data and normalized $P_{abs}(H_{app}, \vartheta_H)$ over the full wave-vector regime studied with a single value for the damping of the SAW excited spin wave, $\Gamma = 0.142 \pm 0.008$, that is independent of both $|\mathbf{q}_\parallel|$ and angle. The precision with which we can quantify the spin-wave damping and its angular and wave-vector dependence via the SAW power absorption measurements is shown in Figure 5.[42] Our extracted value for $\Gamma$ is of the same order as the spin-wave damping values ($\Gamma \sim 0.1$) estimated from simulations in previous SAW work on Ni films.[21] These values of the spin-wave damping are significantly higher than the typical values for the Gilbert damping in Ni ($\alpha_0 \sim 0.048$) as measured by uniform mode resonance. We estimate that the additional damping contribution arising from spin-pumping from our 10 nm Ni film into the 15 nm Pt overlayer is $\alpha_{SP} \sim 0.006$. For this estimation, we have assumed that the real part of the mixing conductance is $g_r^{\uparrow\downarrow} = 2 \cdot 10^{15}$ cm$^{-2}$ for the Ni|Pt interface and $\lambda_s^{Pt} = 1.4$ nm for the spin-diffusion length in Pt[43] and we neglect any enhancement of $\alpha_{SP}$ beyond the macrospin theory associated with the fact that the excitation is a surface spin-wave[44] (justifiable in the limit $d \ll \lambda_{SAW}$). The physical mechanisms responsible for this large spin-wave damping are as yet unclear. We speculate that small magnetostrictive deformations in the Ni film generated by the precessing spin-wave can couple into various elastic modes (i.e., bulk longitudinal and transverse phonons) of the LiNbO$_3$ substrate. This coupling could potentially lead to a large spin-wave damping such as we have extracted from experiment.

## IV. DISCUSSION

In this section, we show that SAW power absorption measurements can be used to perform traveling surface spin-wave spectroscopy and quantitatively measure the anisotropic dipolar spin-spin contributions to the spin-wave dispersion – even at moderate wave-vectors in the low $M_s$ Ni system where these contributions are not particularly strong. The structure of the anisotropic dipole interactions, forming the leading order spin-wave contribution to the dispersion, is embedded within the resonant SAW absorption measurements. We first show that this is the case by demonstrating that the power absorption matches well to an analytical theory for the traveling surface spin-wave including the magnetic dipolar interaction. This is first done for two field scans at different $\vartheta_H$ where the dipolar interactions are expected to be weak and strong and it is shown that the full dipolar theory quantitatively captures the line-shapes of both scans whereas the theory excluding dipolar interactions agrees with the data only where the dipolar corrections to the spin-wave dispersion are expected to be weak. Then we show that the SAW power absorption can be mapped to and sampled over a large part of $(H_{app}, \varphi_0)$ space and agrees quantitatively with the analytical theory including dipolar interactions. Given the quantitative agreement between the power absorption data and the traveling surface dipolar spin-wave theory, we then directly extract the strength of the dipole interactions from the power absorption data at the different $\mathbf{q}_\parallel$ investigated in this study.

We first plot in Figure 6 normalized $|S_{21}^{norm}(H_{app}, \vartheta_H)|$ lineshapes together with the results of calculations for $P_{abs}(H_{app})$ in which both dipolar and exchange interactions are included or in which the dipolar interactions are neglected (*i.e.*, exchange-only), at $\mathbf{q}_\| = 6.9 \times 10^4$ cm$^{-1}$ $\hat{\boldsymbol{\eta}}$ ($f_{pump} = 3.86$ GHz) for two field angles $\vartheta_H = 20°$ and $\vartheta_H = 60°$. The two particular field angles were chosen due to the very different range of $\varphi_0$ sampled in the two field scans. Due to the large in-plane $H_k$, $\varphi_0$ runs through a continuous range of values as $H_{app}$ is swept at some fixed $\vartheta_H$ (except for sweeps on the easy $\zeta$ axis or along directions very close to $\zeta$ where switching events can occur at low $H_{app}$). The $\vartheta_H = 20°$ field scan (*i.e.*, a near hard axis sweep) is such that $\varphi_0$ quickly converges to a large value as $H_{app}$ is increased. For example, $\varphi_0$ is ~60° for the absorption maximum at $H_{app} = 940$ Oe. The $\vartheta_H = 60°$ field scan, on the other hand, is such that $\varphi_0$ is small for all $H_{app}$ and the magnetization lies close to the easy axis. The angle $\varphi_0$ is small, ~17°, for the absorption maximum at $H_{app} = 460$ Oe. The anisotropic dipolar contribution to the lineshape, $2\pi M_s |\mathbf{q}_\|| d \cos^2 \varphi_0$ in Eqn. (10), should thus be small for the $\vartheta_H = 20°$ scan and large for the $\vartheta_H = 60°$ scan. Indeed the two calculations (with and without the dipolar contribution) nearly overlap in the upper panel of Figure 6 for the $\vartheta_H = 20°$ scan and are in good agreement with the $|S_{21}^{norm}(H_{app}, \vartheta_H)|$ data. There is however a large discrepancy between the data and the exchange-only calculation when $\vartheta_H = 60°$, where the anisotropic dipolar correction should be significant, while the $P_{abs}(H_{app})$ calculation including dipolar interactions agrees well with the measured $|S_{21}^{norm}(H_{app}, \vartheta_H)|$ resonance for the $\vartheta_H = 60°$ scan.

This agreement is not limited to the $(H_{app}, \varphi_0)$ space subtended by the $\vartheta_H = 20°$ and $\vartheta_H = 60°$ field scans. Knowledge of $H_k$, $H_{app}$, and $\vartheta_H$ allows for a direct mapping of $|S_{21}^{norm}(H_{app}, \vartheta_H)|$ to $|S_{21}^{norm}(H_{app}, \varphi_0)|$ via the relation $H_k \sin\varphi_0 \cos\varphi_0 = H_{app} \cos(\varphi_0 + \vartheta_H)$, where $\varphi_0$ is the orientation of the magnetization (Figure 1). A plot of $|S_{21}^{norm}(H_{app}, \varphi_0)|$ at $|\mathbf{q}_\parallel| = 6.9 \times 10^4$ cm$^{-1}$ is shown in the top panel of Figure 7. The angular dependence of the dipolar correction in the spin-wave dispersion is reflected in the measured SAW absorption for a broad range of $\varphi_0 \sim 10°$ to $80°$. This can be seen by the good agreement between $|S_{21}^{norm}(H_{app}, \varphi_0)|$ and $P_{abs}(H_{app}, \varphi_0)$ over this entire set of $\varphi_0$ values. The angular range over which SAW absorption due to spin-wave excitation can be accessed is only limited by the fact that the magnetoelastic interaction itself becomes vanishingly small as $\varphi_0$ approaches $0°$ or $90°$. The comparison shown in the bottom panel of Fig. 5 between the measured $|S_{21}^{norm}(H_{app}, \varphi_0)|$ and the isotropic exchange-only calculation shows graphically where and how the anisotropic dipolar correction $2\pi M_s |\mathbf{q}_\parallel| d \cos^2 \varphi_0$ becomes important as a function of $\varphi_0$. As expected, the influence of dipolar interactions on the SAW power absorption are seen to be strong at low $\varphi_0$ and weak for large $\varphi_0$ nearer to $90°$.

The $|S_{21}^{norm}(H_{app}, \varphi_0)|$ data quantitatively match the $P_{abs}(H_{app}, \varphi_0)$ calculation also not just at $|\mathbf{q}_\parallel| = 6.9 \times 10^4$ cm$^{-1}$ but at all the other $\mathbf{q}_\parallel$ employed in the study. We show this by plotting the fields at which maximum SAW absorption occurs at the different $|\mathbf{q}_\parallel|$ allowed by our delay line and at a fixed $\vartheta_H = 45°$ (Figure 8). The positions of the absorption maxima cannot be simultaneously fit for all $|\mathbf{q}_\parallel|$ using the pure-exchange theory and treating $A_{ex}$, $M_s$, and $H_k$ as free parameters or using a uniform-mode resonance theory. A comparison between the pure-exchange theory and the data shows that the corrections to the dispersion beyond pure-exchange model have the wave-vector dependence corresponding to the anisotropic dipolar interaction. We note that previous studies[19,21] operated in a low wave-vector regime ($|\mathbf{q}_\parallel| = 3 \times 10^3 - 4 \times 10^4$ cm$^{-1}$) where the leading order wave-vector dependent contributions beyond the uniform-mode theory are small. These studies model their absorption data using a uniform-mode resonance theory – as is appropriate to most of this low wave-vector range in a Ni film. Our experiment performed at higher wave-vectors clearly measures the impact of the leading order spin-wave contributions to the dispersion (i.e. the dipolar interaction), thus providing a first confirmation that the SAW is resonantly exciting a traveling surface spin-wave matched to the SAW's in-plane wave-vector $\mathbf{q}_\parallel$.

This outstanding quantitative agreement between our measurements and the theory including anisotropic dipolar interactions allows us to use SAW-driven spin-wave spectroscopy to directly extract the strength of the dipole interactions. To do so, we perform a least-squares fit to the $|S_{21}^{norm}(H_{app}, \varphi_0)|$ dataset at various higher $|\mathbf{q}_\parallel|$ where the effects of the dipolar interaction clearly affect the angular dependence of the absorption. We fit the data to the formula (Eqn. (9)) for $P_{abs}(H_{app}, \varphi_0)$, with three free parameters: a constant spin-wave damping $\Gamma$, and the coefficients $A_1$ and $A_2$, defined in Eqn. (12):

$$\Psi = H_k \cos(2\varphi_0) + H_{app} \sin(\varphi_0 + \vartheta_H) + \frac{2A_{ex}}{M_s}|\mathbf{q}_\parallel|^2 + A_1 |\mathbf{q}_\parallel| \cos^2 \varphi_0$$
$$\Upsilon = H_k \cos^2 \varphi_0 + H_{app} \sin(\varphi_0 + \vartheta_H) + \frac{2A_{ex}}{M_s}|\mathbf{q}_\parallel|^2 + \left(4\pi M_s - \frac{2K_\perp}{M_s}\right) + A_2 |\mathbf{q}_\parallel| \quad (12)$$

The analytical theory of Eqn. (5) predicts $A_1 = -A_2 = 4\pi M_s \frac{d}{2} = (3.0 \pm 0.1) \times 10^{-3}$ Oe cm arising from the dipolar spin-wave corrections (with the uncertainty arising from the determination of $M_s$). The best-fit parameters extracted from the data at different $f_{pump}$ and $|\mathbf{q}_\parallel|$ are given in Table 1:

| $f_{pump}$ [GHz] | $|\mathbf{q}_\parallel|$ [$10^4$ cm$^{-1}$] | $\Gamma$ | $A_1$ [$10^{-3}$ Oe cm] | $A_2$ [$10^{-3}$ Oe cm] |
| --- | --- | --- | --- | --- |
| 3.26 | 5.9 | $0.143 \pm 0.006$ | $3.0 \pm 0.2$ | $-(3.1 \pm 0.1)$ |
| 3.86 | 6.9 | $0.139 \pm 0.007$ | $3.1 \pm 0.1$ | $-(2.9 \pm 0.2)$ |
| 4.45 | 8.0 | $0.14 \pm 0.01$ | $2.9 \pm 0.3$ | $-(2.8 \pm 0.3)$ |

Table 1. Coefficients for the dipolar interaction $A_1$ and $A_2$ as well as the spin-wave damping $\Gamma$ extracted from least-squares fitting to $|S_{21}^{norm}(H_{app},\varphi_0)|$ data at various value of $|\mathbf{q}_\parallel|$. Direct extraction of $A_1$ and $A_2$ from the $|S_{21}^{norm}(H_{app},\varphi_0)|$ dataset at $f_{pump} = 2.67$ GHz ($|\mathbf{q}_\parallel| = 4.8 \times 10^4$ cm$^{-1}$) is tricky due to the weaker effect that the dipolar contribution has on the dispersion at this wave-vector and over most of the larger $\varphi_0$ at which $|S_{21}^{norm}(H_{app},\varphi_0)|$ has appreciable amplitude at this $f_{pump}$.

The coefficients $A_1$ and $A_2$ as well as the constraint $A_1 = -A_2$ have been obtained directly from the $|S_{21}^{norm}(H_{app},\varphi_0)|$ measurements at several $|\mathbf{q}_\parallel|$ and agree within an accuracy of a few percent with the expected contributions to the spin wave dispersion arising from the anisotropic dipolar interaction.

## V. CONCLUSION/OUTLOOK

We have quantitatively demonstrated that SAWs can excite a single traveling surface spin-wave mode with an in-plane wave-vector $\mathbf{q}_\parallel$ matched to the wave-vector of the SAW.

Measurements of the SAW power absorption on such traveling surface spin-wave modes can be used to directly extract precise, quantitative information about the angle and wave-vector dependence of the spin-wave dispersion and damping. For the Ni thin films of the present study we find, as expected, that the spin-wave anisotropy is accounted for quantitatively by the dipolar spin-wave interaction. However, this spectroscopic technique can also be implemented to examine other sources of anisotropy. The same technique can be extended to study spin-wave physics that emerges at higher wave vectors (*e.g.*, to map the angular structure of the Dzyaloshinskii-Moriya exchange interaction) by using piezoelectric substrates with ultra-low SAW speeds, initially developed with the objective of obtaining signal delay lines operating at long delay times. For example, wave vectors as large as $|\mathbf{q}_{\parallel}| = 7.9 \times 10^5$ cm$^{-1}$ at 11 GHz could be achieved using SAWs propagating on a (110)-cut Tl$_3$TaSe$_4$ substrate, which have a speed of just $8.9 \times 10^4$ cm/s.[45] This large wave vector is accessible by operating a SAW delay line defined by deep UV lithography at a high-order bandpass (*e.g.*, 11$^{th}$ overtone). Anisotropic Dzyaloshinskii-Moriya contributions to the spin-wave dispersion should be as large as several hundreds of Oe at this wave vector in a system with high interfacial DMI strength (*e.g.*, the Pt|Co system with an interfacial DMI strength of ~ 0.5 ergs/cm$^2$)[5]. Measurements of the angular dependence of $\left|S_{21}^{norm}(H_{app}, \varphi_0)\right|$ would readily allow one to extract the DMI strength. In addition, comparison of two-dimensional $\left|S_{21}^{norm}(H_{app}, \varphi_0)\right|$ plots with $\left|S_{12}^{norm}(H_{app}, \varphi_0)\right|$ plots would allow for a direct measure of the spin-wave non-reciprocity ($\mathbf{q}_{\parallel} \to -\mathbf{q}_{\parallel}$) associated with the DMI.

    A SAW-based spectroscope should also be an excellent tool for carrying out wave-vector and magnetization-angle-resolved studies on 1D and 2D magnonic crystals composed of ultra-thin magnetic elements. In such a case, SAW modification associated with periodic reflections

off the magnonic crystal (arising from *e.g.,* mass loading) should be very small, providing SAW based spectroscopy the capability to map spin-wave dynamics throughout the magnonic Brillouin zone.

## **Acknowledgements**

We acknowledge G.E. Rowlands for help with Figure 1 of the manuscript. We also thank S. Aradhya and G. Finnochio for suggestions on the manuscript. We are grateful to T. Gosavi and S. Bhave for their encouragement throughout the course of the project. This work was supported in part by the Office of Naval Research and the Army Research Office. This work made use of the Cornell Center for Materials Research Shared Facilities, which are supported through the NSF MRSEC program (DMR-1120296). This work was performed in part at the Cornell NanoScale Facility, a node of the National Nanotechnology Infrastructure Network, which is supported by the National Science Foundation (Grant ECCS-0335765).

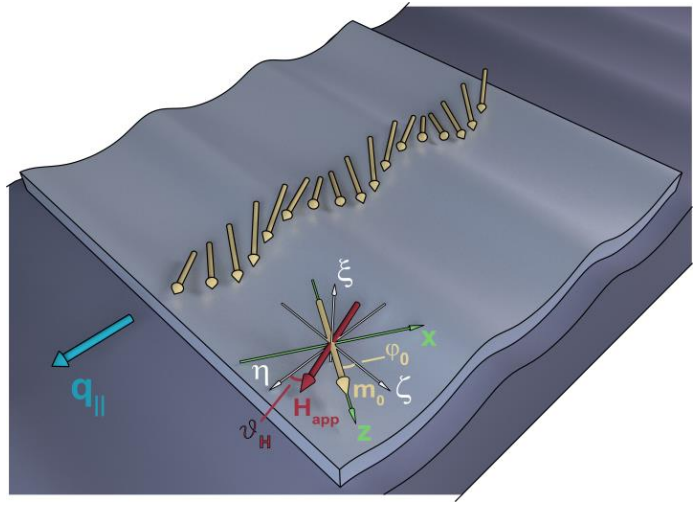

Figure 1. Illustration of a traveling SAW with wave vector $\mathbf{q}_\parallel$ generating a time-dependent traveling strain wave field in a Ni film and driving a spin-wave resonance. This figure defines the $\zeta\eta\xi$ coordinate system, the equilibrium magnetization $\mathbf{m_0}$ and angle $\varphi_0$, the $xyz$ coordinate system used for linearizing the LLG equations about equilibrium, and the in-plane applied magnetic field and field angle $\vartheta_H$.

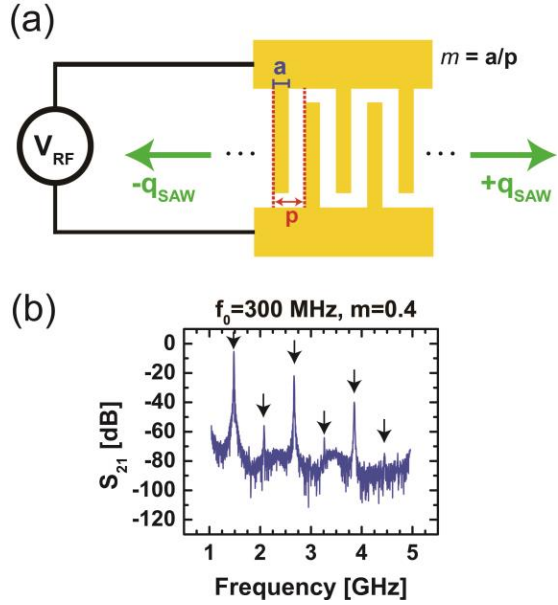

Figure 2. a) Illustration of inter-digitated electrodes used for launching a SAW. The signal and ground finger pattern is repeated to produce the full IDT electrode used in our study with $N$ = 40 total fingers. We have used electrodes with pitch $p = \lambda_0 / 2 = 5.8$ μm and metallization ratio $m = a / p = 0.4$, where $\lambda_0$ is the fundamental band pass center wavelength and $a$ is the IDT finger width. The wave vector of the SAW launched by the IDT is $\mathbf{q}_{SAW} \sim \mathbf{q}_\parallel$.[46] b) Time-gated $S_{21}$ transmission spectrum for our fundamental $f_0 = 300$ MHz SAW delay line where emitter

and receiver electrodes are placed at a center-to-center distance of 700 μm. High-order bandpass center frequencies are visible at $f_{pump}$ = 1.48 GHz, 2.08 GHz, 2.67 GHz, 3.26 GHz, 3.86 GHz, and 4.45 GHz. The gate center time for the spectrum was set at ~ 0.2 μs in order to maximize the single-transit signal, and we used a gate span of 0.05 μs. All of our SAW-driven resonance measurements used a time-gate with this set of specifications.

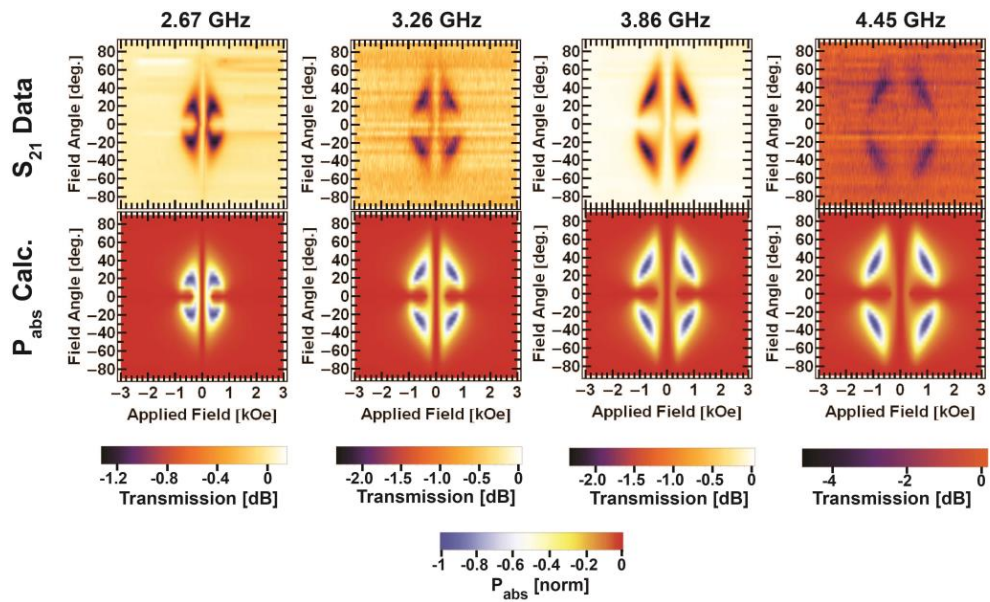

Figure 3. Simultaneous plots of the measured log scale $\left|S_{21}^{loss}(H_{app},\vartheta_H)\right|$ transmission loss and the normalized $P_{abs}$ calculation as a function of field angle $\vartheta_H$ and field magnitude $H_{app}$.

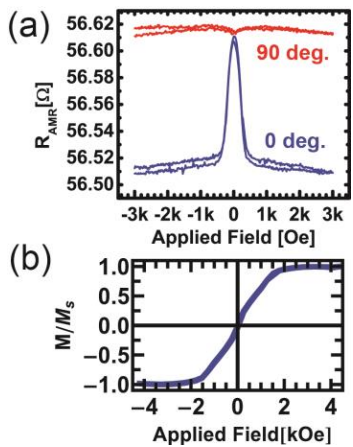

Figure 4. a) AMR curves for 0° (blue) and 90° (red) field angle with respect to the SAW propagation direction show that the Ni has a substantial in-plane anisotropy field $H_k$ with 90°

(*i.e.*, the $\zeta$ axis) being the easy-axis direction. b) Out-of-plane field scan of Al(10)/AlO$_x$(2)/Ni(10)/Pt(15) bilayer on YZ-cut LiNbO$_3$.

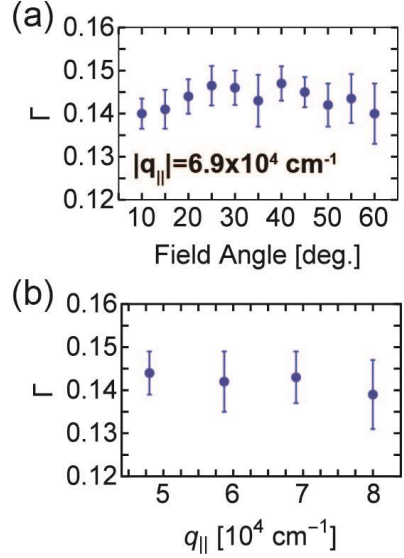

Figure 5. a) Spin wave damping measured as a function of $\vartheta_H$ at fixed $|\mathbf{q}_\parallel| = 6.9 \times 10^4$ cm$^{-1}$. Scans at other wave vectors show similar angle-independent damping. Error bars indicate the standard error as obtained from least-squares fits to normalized, linear power $|S_{21}^{norm}(H_{app}, \vartheta_H)|$ lineshapes. b) Spin-wave damping averaged over field angle as a function of wave vector shows that spin-wave damping is wave-vector independent. Error bars for damping as a function of $|\mathbf{q}_\parallel|$ indicate standard error obtained from least squares fits to the series of $\vartheta_H$ scans (at fixed $|\mathbf{q}_\parallel|$) for which there is appreciable signal.

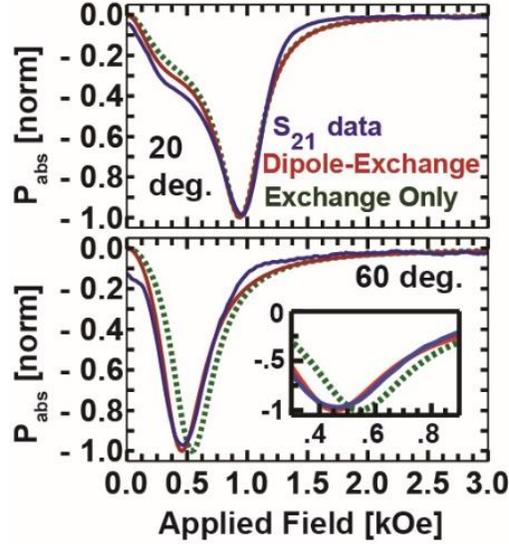

Figure 6. Comparison of normalized $\left|S_{21}^{norm}(H_{app},\vartheta_H)\right|$ transmission data with the power absorption $P_{abs}$ predicted by an exchange-only theory and also a full theory including dipolar corrections, for $\vartheta_H = 20°$ and $60°$ at $f_{pump} = 3.86$ GHz ($|\mathbf{q}_\parallel| = 6.9\times10^4$ cm$^{-1}$). The data normalization has been carried out by converting the $S_{21}$ loss to linear power and rescaling each $S_{21}$ vs $H_{app}$ curve to lie between -1 and 0.

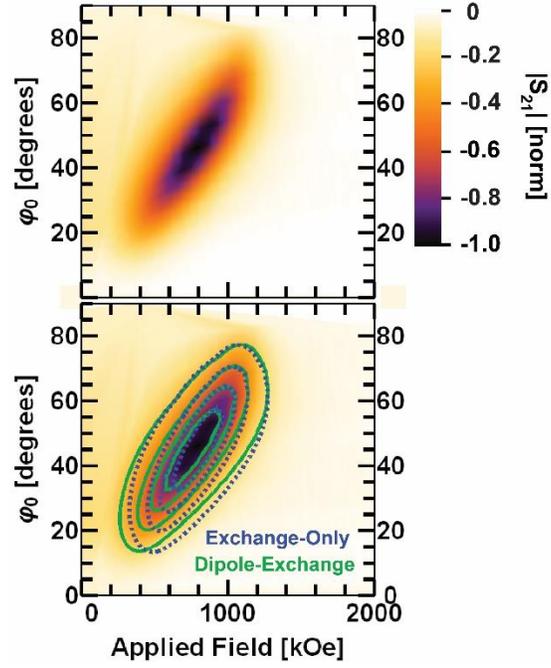

Figure 7. Plot of the normalized $|S_{21}^{norm}(H_{app},\varphi_0)|$ at $|\mathbf{q}_\parallel| = 6.9 \times 10^4$ cm$^{-1}$. The bottom panel has the contours of the dipole-corrected $|P_{abs}(H_{app},\varphi_0)|$ and the exchange-only calculation overlaid on top of the $|S_{21}^{norm}(H_{app},\varphi_0)|$ data. The contours for each calculation vary in equal steps from $P_{abs} = -0.8$ (innermost) to $P_{abs} = -0.2$ (outermost).

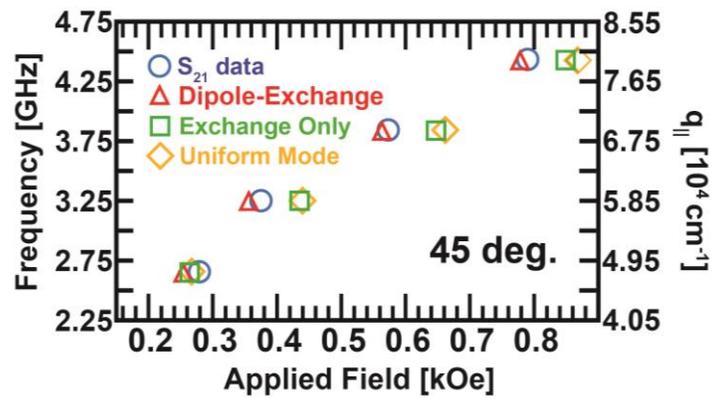

Figure 8. Comparison of the field values where the absorption maximum should occur as a function of pump frequency for $\vartheta_H = 45°$ scans as predicted by the uniform mode resonance theory, an exchange-only theory and a full surface dipole-exchange theory, and as measured from the SAW absorption.

[37] Our analysis of SAW-driven spin-wave resonance treats the elastic degrees of freedom in the Al/AlO$_x$/Ni/Pt multilayer as determined by the propagating SAW in the LiNbO$_3$. In this case, the influence of the strain on the magnetic system can be described in terms of an effective magnetic field in the LLG equation. This assumption is valid when the film thickness is considerably smaller than the SAW wavelength $\lambda_{SAW}$ (which roughly corresponds to the penetration depth of the SAW into the bulk of the solid). In this limit the strain in the thin film can be considered, to a good approximation, as possessing the same profile and behavior of the SAW at the top boundary of the LiNbO$_3$. The film, in this regime, alters slightly the electromechanical boundary condition on the SAW, and generates a weak renormalization of the SAW velocity and SAW absorption. The effective boundary condition produced by the film stack also includes the effect of the magnetoelastic interaction in the Ni film responsible for spin-wave resonance.

[42] We note that the fitting of $\left|S_{21}^{norm}(H_{app},\vartheta_H)\right|$ using the theoretical expression for $P_{abs}$ does not include contributions to the lineshape arising from inhomogenous broadening. The fact that the line-shapes at different $f_{pump}$ and $|\mathbf{q}_\parallel|$ are well fit by a single spin-wave damping $\Gamma$ seems to indicate that this contribution is small and that the broadening is spin-wave damping dominated at the different $f_{pump}$ and $H_{app}$ in this experiment. This will not be true in general for low damping systems pumped at low $f_{pump}$ and Eqn. (9) will need to be modified to account for inhomogenous broadening.

[46] We note that the in-plane wave vector $\mathbf{q}_{SAW}$ of the SAW launched at the IDT at a given pump frequency is not exactly the same as the $\mathbf{q}_\parallel$ of the SAW traveling in and under the Al/AlO$_x$/Ni/Pt thin film. The difference between the two arises from the fact that $c_{SAW}$ changes as the SAW propagates under the thin film stack. We estimate that this change in velocity change (and corresponding change in $\mathbf{q}_\parallel$) is small and on the order of 2% with the dominant contribution coming from the capacitive coupling of the SAW to the metallic film stack. Changes in $\mathbf{q}_\parallel$ on the order of 2% have a negligible impact on the analysis and fitting of the absorption lineshapes given uncertainties in $H_k$ and $M_s$. We therefore have assumed that $|\mathbf{q}_\parallel| \approx |\mathbf{q}_{SAW}|$ in the main text.